\documentclass[11pt,a4paper]{article}
\usepackage{amsmath,amssymb}
\newcommand{\JPA}[3]{J.~Phys.\ A {\bfseries #1}, #2 (#3)}
\newcommand{\PRL}[3]{Phys.\ Rev.\ Lett.\ {\bfseries #1}, #2 (#3)}
\newcommand{\PRB}[3]{Phys.\ Rev.\ B {\bfseries #1}, #2 (#3)}
\newcommand{\JPC}[3]{J.~Phys.~C {\bfseries #1}, #2 (#3)}

\newcommand{\JPhysI}[3]{J.~Phys.~I France {\bfseries #1}, #2 (#3)}

\newcommand{\EPJB}[3]{Eur.~Phys.~J.~B {\bfseries #1}, #2 (#3)}

\newcommand{\IK}{I.~Kondor}
\newcommand{\CD}{C.~De Dominicis}
\newcommand{\TT}{T.~Temesv\'ari}

\begin{document}

\title{Generic replica symmetric field-theory for short range Ising
       spin glasses}

\author{
        T.~Temesv\'ari\thanks{Corresponding author\newline Address:
        Institute for Theoretical Physics, E\"otv\"os University,
        H-1518 Budapest, Pf.~32, Hungary\newline 
        E-mail: temtam@helios.elte.hu\newline
        Telephone: (36)-(1)-2090555-6126\newline
        Fax: (36)-(1)-3722509}
        \\HAS Research Group for Theoretical Physics,
        E\"otv\"os University,\\H-1117 P\'azm\'any P\'eter s\'et\'any 1/A,
        Budapest, Hungary
\and
        C.~De Dominicis\\Service de Physique Th\'eorique,
        CEA Saclay,\\F-91191 Gif sur Yvette, France
\and
        I.R.~Pimentel\\Department of Physics and CFMC, University
        of Lisbon,\\1649 Lisboa, Portugal
       }

\maketitle

\begin{abstract}
Symmetry considerations and a direct, Hubbard-Stratonovich type, derivation are
used to construct a replica field-theory relevant to the study of the spin
glass transition of short range models in a magnetic field. A mean-field
treatment reveals that two different types of transitions exist,
whenever the replica number $n$ is kept larger than zero.
The Sherrington-Kirkpatrick critical point in zero  magnetic field
between the paramagnet and replica
magnet (a replica symmetric phase with a nonzero spin glass order parameter)
separates from the de Almeida-Thouless line,
along which replica symmetry breaking occurs.
We argue that
for studying the de Almeida-Thouless transition around the upper critical
dimension $d=6$, it is necessary to use the generic cubic model with all
the three bare masses and eight cubic couplings. The critical role $n$ may
play is also emphasized. To make perturbative calculations feasible, a new
representation of the cubic interaction is introduced. To illustrate
the method, we compute the masses in one-loop order. Some technical details
and a list of vertex rules are presented to help future renormalisation-group
calculations.

\end{abstract}

\section{Introduction}

The Ising spin glass is the simplest model still incorporating all the complexity
that
more sophisticated disordered systems show up. As such, it has become widely studied
in the last decades.
We focus our attention here to the case where Ising
spins interact via Gaussian-distributed pair interactions 
\cite{review}. A huge amount of
literature has accumulated since the seminal paper of Edwards and Anderson
\cite{EA}, nevertheless the most important problems---i.e.\ the nature and complexity
of the glassy phase, the existence of a transition in nonzero magnetic field,
temperature-chaos, etc.---are still debated. Consensus has been reached only
for mean-field theory, first derived considering an infinite number of fully-connected
Ising spins \cite{SK}; its solution by Parisi explicitly breaks
the replica symmetry, resulting in a picture where the glassy phase can be
decomposed into a set of infinite number and ultrametrically organised
pure states \cite{review}. Despite all efforts made to go beyond
mean-field theory \cite{beyond},
which is certainly valid in infinite dimensions,
finite-dimensional short-ranged systems are much less
understood. Beside the mean-field picture, an alternative description,
the so-called ``droplet picture'' has emerged in a series of papers
(a list of them, which is certainly not fully complete, is provided as
Ref.\ \cite{droplet}). This approach claims that replica symmetry breaking
is an artifact of mean-field theory, and the glassy phase consists of only
two pure phases related by a global inversion of the spins. The droplet theory
has gained some support from the field of mathematical physics \cite{NewmanStein},
the conclusions, however, remain disputed \cite{antiNewmanStein,numeric,RSB}. 

A large amount
of numerical work\footnote{An extensive list of references for numerical
simulations in spin glasses
can be found in \cite{numeric}.} has been carried out to resolve the problem,
finite-size effects and long relaxation times, however, make it difficult to
reach a definite conclusion. It is clear that analytical methods---especially
field-theory, as the most powerful of them---are very important to
provide reliable results to settle this controversy. A direct field-theoretical
study of the glassy phase, however, has proved to be very hard, due to the
complexity of the Gaussian propagators and the ubiquity of infrared divergences
(see \cite{beyond} and references therein). A scaling theory for the spin glass
phase (just below $T_c$ and in zero magnetic field) and a proposal to handle
the infrared problems were put forward in Ref.\ \cite{infrared},
still progress in that direction is very slow.

There is one characteristic of the phase diagram in the mean-field picture
which is definitely absent in a droplet-like approach, namely the
existence of a spin glass transition in a uniform magnetic field
\cite{AT}, known as the de Almeida-Thouless (AT) transition.
This question can be studied in the {\em symmetric} phase,
approaching the presumed transition from the high-temperature side, in
this way eliminating the problems arising from the complexity of the
glassy phase. In the language of replica field-theory, this will lead to
a replica symmetric Lagrangean which is invariant under any permutation of
the $n$ replicas.

The prime purpose of this paper is to provide the generic field-theoretical
model appropriate to a detailed study of the problem raised above, i.e.\ the
existence of an AT transition. (In a separate publication
\cite{Iveta}, the crossover region around the zero-field critical
point is elaborated, and the role that a small magnetic field plays
in driving the AT transition is
investigated.) Here we rediscover, at the mean-field level,
the importance of the replica number $n$ in the 
analysis of the AT transition \cite{Kondor,PCS}: for $n$ small but nonvanishing,
the AT transition line moves away from the zero-field critical
point, and an intermediate range of temperature emerges even in zero
magnetic field. This phase---which can be called, by the extension of the
concept of Sherrington \cite{Sherrington} to continuous $n$, the replica
magnet phase---is replica symmetric but still has a lower symmetry than
the paramagnetic phase. Hence we have two transitions, the first one,
in zero field, is
an isolated critical point between the two replica symmetric phases
(paramagnet and replica magnet), whereas the second one is a whole line
in $H-T$ plane between replica magnet and the replica symmetry broken
phase. As a result, we can identify the AT transition as the onset of
instability of the replica magnet phase, and since it has a lower symmetry
than the paramagnet, we must use a generic replica symmetric Lagrangean
to study it by field-theoretical methods.\footnote{We keep $n\gtrsim 0$ small
but finite almost everywhere throughout the paper. This is because we want to
present formulae for later calculations in the generic cubic model. For this,
however, the $n\to 0$ limit is rather tricky, due to the degeneracy of the
longitudinal and anomalous modes at zero $n$.}

From Eqs.\ (\ref{L2}) and (\ref{L3}), one can immediately
realize that a perturbational calculation based on that Lagrangean is
extremely difficult.
This is due to the complicated interaction term with
eight different couplings corresponding to the eight possible cubic
invariants, and also to the non-diagonal Gaussian-propagators with
three distinct bare masses. To overcome these difficulties, we
introduce a new representation of the cubic interaction which is associated
with a
block-diagonalization of the quadratic part. The technique proves to be very
efficient, as is displayed in our example where the one-loop calculation
of the mass operator is presented.%
\footnote{Other methods are also available, like Replica
Fourier Transform \cite{RFT}, or the usage of projections to the
subspaces of the fundamental modes \cite{Iveta,BrRo}.}

The outline of the paper is as follows: In Sec.\ \ref{cubic} the generic
cubic Lagrangean for the replica symmetric field-theory is set forth,
first using only symmetry considerations. It is then {\em derived},
starting from the lattice system, and using the standard Hubbard-Stratonovich
transformation.
Sec.\ \ref{analysis} is devoted to an analysis of the zero-loop,
i.e.\ mean-field, results. The zero-field transition, first discovered by
Sherrington and Kirkpatrick \cite{SK}, proves to be an isolated
singularity of the stationary condition, with the unique mass vanishing
at that point. On the other hand, one of the three modes becomes
massless along the AT surface, signalling the onset of instability of
the generic replica symmetric phase. In Sec.\ \ref{canonical} we define
the new set of cubic couplings. The introduction of this new representation
makes it possible to compute Feynmann-graphs of a perturbative approach;
this is illustrated in a one-loop calculation of the three masses in
Sec.\ \ref{illustration}. A simple and convenient non-orthogonal basis is
presented in \appendixname\ A, whereas a detailed list of vertices computed
in this basis is given in \appendixname\ B. This almost complete table of
vertex rules is published here for later references, an application for an
extended renormalisation group study of the finite-dimensional AT
transition is in progress \cite{letter}.

\section{Cubic replica field-theory for the Ising spin glass in nonzero
magnetic field}\label{cubic}

After the invention of the renormalization group \cite{Wilson},
field theoretical representations of statistical models, originally defined
on a lattice, became a standard way to study the behaviour
of the systems near phase
transitions. 
The renormalisation group made it possible to use a perturbative method, the
loop expansion, in low enough dimensions, thus
providing excellent analytical tools to compute phase diagrams and critical
properties. The extension to spin glasses came immediately after
the replica approach had been introduced by Edwards and Anderson \cite{EA},
transforming the originally inhomogeneous system into a homogeneous one.
A Ginzburg-Landau-Wilson continuum model was first proposed 25
years ago \cite{Harris,CheLu},
then further investigated by renormalisation methods
\cite{Elderfield,PR}. Its cubic Lagrangean was derived---for
the Ising case---by Bray and Moore \cite{BrMo}
via the Hubbard-Stratonovich transformation.
Two of the present authors applied
the same field theoretical model in their effort to go beyond the
mean-field results, and understand the glassy phase of the
finite-dimensional short range Ising spin glass \cite{beyond,infrared}.
The magnetic field was always zero in the above works, with the only
exception of \cite{CheLu}, where it was introduced by a coupling to
the magnetization, leaving the Lagrangean unchanged for the part relevant
to the spin glass transition.

Field-theoretical models can be constructed by means of symmetry arguments,
building up the Lagrangean from
all the possible invariants of the relevant symmetry group
of the system. For an Ising spin glass, the fields depend on a pair of
replicas, $
\phi^{\alpha\beta}
=\phi^{\beta\alpha}$ with
$\phi^{\alpha\alpha}=0$, and, as a consequence of the replica trick,
any permutation of the $n$ replicas leaves the Lagrangean unchanged.
Discarding all the terms higher in the order of the $\phi$'s than cubic,
we arrive at the following {\em generic replica symmetric} Lagrangean
after a search of all the quadratic and cubic invariants:
\[ \mathcal{L}=\mathcal{L}^{(2)}+\mathcal{L}^{(3)},\] where
\begin{multline}\label{L2}
\mathcal{L}^{(2)}=\frac{1}{2}\sum_{\mathbf p}\bigg[
\Big(\frac{1}{2} (pa\rho)^2+m_1\Big)\sum_{\alpha\beta}
\phi^{\alpha\beta}_{\mathbf p}\phi^{\alpha\beta}_{-\mathbf p}
+m_2\sum_{\alpha\beta\gamma}
\phi^{\alpha\gamma}_{\mathbf p}\phi^{\beta\gamma}_{-\mathbf p}\\+
m_3\sum_{\alpha\beta\gamma\delta}
\phi^{\alpha\beta}
_{\mathbf p}\phi^{\gamma\delta}_{-\mathbf p}
\bigg],
\end{multline}
and
\begin{multline}\label{L3}
\mathcal{L}^{(3)}=-\frac{1}{6\sqrt{N}}
\sideset{}{'}\sum_{\mathbf {p_1p_2p_3}}
\bigg[
w_1\sum_{\alpha\beta\gamma}\phi^{\alpha\beta}
_{\mathbf p_1}\phi^{\beta\gamma}_{\mathbf p_2}
\phi^{\gamma\alpha}_{\mathbf p_3}+w_2\sum_{\alpha\beta}\phi^{\alpha\beta}
_{\mathbf p_1}\phi^{\alpha\beta}
_{\mathbf p_2}\phi^{\alpha\beta}_{\mathbf p_3}\\[5pt]
+w_3\sum_{\alpha\beta\gamma}\phi^{\alpha\beta}
_{\mathbf p_1}\phi^{\alpha\beta}
_{\mathbf p_2}\phi^{\alpha\gamma}_{\mathbf p_3}
+w_4\sum_{\alpha\beta\gamma\delta}\phi^{\alpha\beta}
_{\mathbf p_1}\phi^{\alpha\beta}
_{\mathbf p_2}\phi^{\gamma\delta}_{\mathbf p_3}
+w_5\sum_{\alpha\beta\gamma\delta}\phi^{\alpha\beta}
_{\mathbf p_1}\phi^{\alpha\gamma}
_{\mathbf p_2}\phi^{\beta\delta}_{\mathbf p_3}\\[2pt]
+w_6\sum_{\alpha\beta\gamma\delta}\phi^{\alpha\beta}
_{\mathbf p_1}\phi^{\alpha\gamma}
_{\mathbf p_2}\phi^{\alpha\delta}_{\mathbf p_3}
+w_7\sum_{\alpha\beta\gamma\delta\mu}\phi^{\alpha\gamma}
_{\mathbf p_1}\phi^{\beta\gamma}
_{\mathbf p_2}\phi^{\delta\mu}_{\mathbf p_3}
+w_8\sum_{\alpha\beta\gamma\delta\mu\nu}\phi^{\alpha\beta}
_{\mathbf p_1}\phi^{\gamma\delta}
_{\mathbf p_2}\phi^{\mu\nu}_{\mathbf p_3}
\bigg].
\end{multline}
Momentum summations in the above formulae are over the reciprocal
vectors of a $d$-dimensional hypercubic lattice with lattice
spacing $a$, consisting of infinitely many sites $N$ in the
thermodynamic limit. The prime in Eq.\ (\ref{L3})
means the constraint of
momentum conservation, $\mathbf p_1+\mathbf p_2+
\mathbf p_3=0$. Neglecting the fluctuations of fields with wavelength
much smaller than the range $\rho a$
of the exchange interaction between the spins,
we confine the relevant values of momentums in
Eqs.~(\ref{L2},\ref{L3}) around the center of the Brillouin zone:
$p< \Lambda /\rho a$. The momentum cutoff $\Lambda \ll 1$ makes it possible to
expand the nonlocal quadratic coupling in ${\mathcal L}^{(2)}$, and, as it
is common in field-theoretical studies of phase transitions, we stop after
the first two terms. 
(The coupling constant of
the kinetic term in
Eq.~(\ref{L2}) can be set equal to $\frac{1}{2}$
without loss of generality. See also later.)
\footnote{A replica symmetric treatment of the {\em ordered\/} phase
was carried out in Refs.~\cite{PR,BrMo}. In this case, the finite
order parameter gives
rise to a quadratic Lagrangean, even in zero field, which is a special
case of ${\mathcal L}^{(2)}$.}

In zero magnetic field, and in the high-temperature phase
where the spin glass order parameter is zero,
all the couplings but $m_1$ and $w_1$ disappear.
In this section,
we want to find out the order parameter dependence of the couplings defining our
Lagrangean. We are especially interested in
the general form of the replica field-theory
suitable for studying the de Almeida-Thouless transition in finite
dimensions. Our starting point is a standard Edwards-Anderson-like
\cite{EA} model for $N$ Ising spins on a $d$-dimensional hypercubic lattice,
with a long but finite-ranged interaction:
\begin{equation}\label{EAHamiltonian}
\mathcal{H}=-\sum_{(ij)}\frac{J_{ij}}{\sqrt{z}}f_{ij} s_is_j 
-H\sum_i s_i.
\end{equation}
The notation $f_{ij}\equiv f\left(\frac{|\mathbf r_i-\mathbf r_j|}
{\rho a}\right)$ was introduced in the above equation, with the \mbox{smooth}
positive function $f(x)$ which takes the value $1$ for $x\lesssim 1$,
and decays to zero sufficiently fast for $x>1$, thereby cutting off the
interaction around $|\mathbf r_i-\mathbf r_j|\sim \rho a$.
Here $z=\rho^d$
is effectively the coordination number, i.e.\ the number of spins within
the interaction radius; expanding quantities in
terms of its negative powers will generate the loop-expansion in the
replica field-theory. $J_{ij}$ are independent, Gaussian distributed
random variables with mean zero and variance $\Delta ^2$,
and a homogeneous magnetic field $H$ was also included. Summations are
over the pairs $(ij)$ of lattice sites in the first sum, while over
the $N$ lattice sites in the second one.

In the spirit of the replica trick, we want to compute quantities like
the averaged replicated partition function $\overline{Z^n}$ for some
positive integer $n$, finally deducing spin glass behaviour from the
$n\to 0$ continuum limit. Averaging first over the quenched disorder
results in an effective replica Hamiltonian depending on the spins
$S_i^{\alpha}$, $\alpha=1,\ldots,n$:%
\footnote{Throughout the paper, we use different notations for
summations over distinct pairs, $\sum\limits_{(\alpha\beta)}=\sum\limits_{\alpha<\beta}$,
and unrestricted sums, $\sum\limits_{\alpha\beta}=
\sum\limits_{\alpha}\sum\limits_{\beta}$.}
\[
\overline{Z^n}\sim \underset{\{S_i^{\alpha}\}}{\mathrm {Tr}}
\exp \bigg(\frac{1}{2}\sum_{(\alpha\beta)}\sum_{ij}S_i^{\alpha}S_i^{\beta}K_{ij}
S_j^{\alpha}S_j^{\beta}+\frac{H}{kT}\sum_\alpha\sum_iS_i^{\alpha}\bigg),
\]
$K_{ij}\equiv\frac{1}{z}\left(\frac{\Delta}{kT}\right)^2f_{ij}^2$.
A Hubbard-Stratonovich transformation can help us to get rid of the
four-spin interaction term; the price we have to pay for that is the
introduction of integrals over the ``fields'' $Q_i^{\alpha\beta}$:
\begin{equation}\label{Qint}
\overline{Z^n}\sim \bigg[\prod_{  (\alpha\beta),
                                     i }\int
dQ_i^{\alpha\beta}\bigg]\exp \bigg(-\frac{1}{2}\sum_{(\alpha\beta)}
\mathbf{Q}^{\alpha\beta}\mathbf{K}^{-1}\mathbf{Q}^{\alpha\beta}+
\sum_i\ln \zeta_i\bigg).
\end{equation} 
The boldfaced vector and matrix notations in Eq.~(\ref{Qint}) for
$\mathbf{Q}^{\alpha\beta}$ and $\mathbf{K}^{-1}$, respectively, are to
simplify the nonlocal term in the formula, whereas the one-spin partition
function is defined as follows:
\begin{equation}\label{zeta}
\zeta_i=\underset{\{S^{\alpha}\}}{\mathrm {Tr}}
\exp \bigg(\sum_{(\alpha\beta)}Q_i^{\alpha\beta}\,S^{\alpha}S^{\beta}+
\frac{H}{kT}\sum_\alpha S^{\alpha}\bigg).
\end{equation}
To construct a field-theory appropriate for a perturbation expansion
around the mean-field solution, i.e.\ around the infinite range model
$\rho \to \infty$, we separate $Q_i^{\alpha\beta}$ into its homogeneous,
non-fluctuating (mean-field) part, and into its fluctuating part:
\begin{equation}\label{Qsep}
Q_i^{\alpha\beta}=Q^{\alpha\beta}+\phi_i^{\alpha\beta}.
\end{equation}
When expressed in terms of the $\phi$'s, the exponent in Eq.(\ref{Qint})
will be called $-\mathcal{L}$, and it can be expanded up to any desired
order. Turning to a more convenient representation of the fields in
momentum space, contributions up to cubic order have the following  forms:
\begin{align}
\label{L0} \mathcal{L}^{(0)} &=N\,\bigg[\frac{1}{2}\Theta^{-1} 
\sum_{(\alpha\beta)}\left(Q^{\alpha\beta}\right)^2-\ln \zeta\bigg],\\
\label{L1} \mathcal{L}^{(1)} &=\sqrt{N}\,
\sum_{(\alpha\beta)}\bigg[\Theta^{-1}Q^{\alpha\beta}-
\langle\langle S^{\alpha}S^{\beta}\rangle\rangle\bigg]
\phi^{\alpha\beta}
_{\mathbf{p}=0}\,,\\
\label{L2M}\mathcal{L}^{(2)} &=\frac{1}{2}\sum_{\mathbf{p}}
\sum_{(\alpha\beta),(\gamma\delta)}\phi^{\alpha\beta}_{\mathbf{p}}
M_{\alpha\beta,\gamma\delta}(pa\rho)\phi^{\gamma\delta}_{\mathbf{-p}},
\\
\intertext{and finally}
\label{L3W}\mathcal{L}^{(3)} &=-\frac{1}{6\sqrt N}
\sideset{}{'}\sum_{\mathbf {p_1p_2p_3}}
\sum_{(\alpha\beta),(\gamma\delta),(\mu\nu)}
W_{\alpha\beta,\gamma\delta,\mu\nu}\phi^{\alpha\beta}_{\mathbf{p}_1}
\phi^{\gamma\delta}_{\mathbf{p}_2}\phi^{\mu\nu}_{\mathbf{p}_3}.
\end{align}
A Boltzmann-weight with $Q^{\alpha\beta}$, instead of
$Q^{\alpha\beta}_i$, is understood in the definitions of $\zeta$ and
the one-site effective average $\langle\langle\ldots\rangle\rangle$
in Eqs.~(\ref{L0},\ref{L1}).
$\Theta^{-1}$ is essentially the temperature
squared, or more precisely:
\begin{equation}\label{Theta}
\Theta=
\left(\frac{\Delta}{kT}\right)^2\int f(r)^2\,d^dr.
\end{equation}
The momentum-{\em dependent\/} mass, and
the momentum-{\em independent\/} cubic coupling operators are defined
as follows:
\begin{align}
\label{M} M_{\alpha\beta,\gamma\delta}(pa\rho)&=K_{\mathbf p}^{-1}
\delta_{\alpha\beta,\gamma\delta}^{\mathrm{Kr}}-\big[\langle\langle
S^{\alpha}S^{\beta}S^{\gamma}S^{\delta}\rangle\rangle -
\langle\langle S^{\alpha}S^{\beta}\rangle\rangle
\langle\langle S^{\gamma}S^{\delta}\rangle\rangle\big],\\[5pt]
W_{\alpha\beta,\gamma\delta,\mu\nu}&=\langle\langle
S^{\alpha}S^{\beta}S^{\gamma}S^{\delta}S^{\mu}S^{\nu}\rangle\rangle
-\langle\langle S^{\alpha}S^{\beta}\rangle\rangle
\langle\langle S^{\gamma}S^{\delta}S^{\mu}S^{\nu}\rangle\rangle
\notag\\[2pt] \label{W}
&-\langle\langle S^{\gamma}S^{\delta}\rangle\rangle
\langle\langle S^{\alpha}S^{\beta}S^{\mu}S^{\nu}\rangle\rangle
-\langle\langle S^{\mu}S^{\nu}\rangle\rangle
\langle\langle S^{\alpha}S^{\beta}S^{\gamma}S^{\delta}\rangle\rangle
\\[2pt]
&+2\langle\langle S^{\alpha}S^{\beta}\rangle\rangle
\langle\langle S^{\gamma}S^{\delta}\rangle\rangle
\langle\langle S^{\mu}S^{\nu}\rangle\rangle .\notag
\end{align}
The Kronecker delta in Eq.~(\ref{M}) represents the $n(n-1)/2$-dimensional
unit matrix, whose prefactor comes from the Fourier-transform of
$K_{ij}$:
\begin{multline}\label{Kp}
K_{\mathbf p}=\frac{1}{N}\sum_{ij}e^{i\mathbf p(\mathbf r_i-\mathbf r_j)}
K_{ij}=\\ \frac{1}{z}\left(\frac{\Delta}{kT}\right)^2
\frac{1}{N}\sum_{ij}e^{i\mathbf p(\mathbf r_i-\mathbf r_j)}
f\left(\frac{|\mathbf r_i-\mathbf r_j|}{\rho a}\right)^2\longrightarrow
\left(\frac{\Delta}{kT}\right)^2 \int e^{i(\mathbf p a\rho)\mathbf r}
f(r)^2\,d^dr.
\end{multline}
The arrow in the above formula means the double limiting procedure
of the thermodynamic limit ($N\to \infty$), and continuum limit
($a\to 0$). The theory resulting then is a field-theory with all
the thermodynamic functions scaling correctly with $N$, and the
lattice constant $a$ disappearing from the momentum integrals
after introducing the new variable $\tilde {\mathbf p}\equiv
\mathbf p a\rho$, with the upper momentum-cutoff
becoming $\Lambda$ in $\tilde {\mathbf p}$
space.
The range $\rho$ of the original interaction, however,
survives:
a perturbative loop-expansion can be generated where
every loop in a Feynmann-diagram contributes a $z^{-1}=\rho^{-d}$
factor.  

It is rather common in field-theoretical studies to normalize the
fields such that the kinetic term in the Gaussian part of the
Lagrangean (that proportional to the squared momentum) be exactly
$\tilde {\mathbf p}^2$ times the unit matrix. This can be simply
reached after expanding $K_{\mathbf p}^{-1}$, Eq.~(\ref{Kp}),
and introducing the new fields by
\begin{equation}\label{newfields}
c\,\phi_{\mathbf p}\rightarrow \phi_{\mathbf p},
\end{equation} 
where
\begin{equation}\label{c}
c=(2d)^{-\frac{1}{2}}\,
\frac{\left(\int r^2f(r)^2\,d^dr\right)^{\frac{1}{2}}}
{\int f(r)^2\,d^dr}\,\left(\frac{kT}{\Delta}\right).
\end{equation}
A corresponding redefinition of the mass operator and cubic
interaction,
\begin{equation}\label{redef}
\frac{1}{c^2}\,M\rightarrow M\quad\text{and}\quad
\frac{1}{c^3}\,W\rightarrow W,
\end{equation}
leaves the form of Eqs.~(\ref{L2M},\ref{L3W}) unchanged.
Neglecting short-wavelength fluctuations, the mass-operator in
Eq.~(\ref{M}) can be expanded for $\tilde p\ll 1$. The commonly used
truncation at the kinetic term provides:
\begin{align}
M_{\alpha\beta,\gamma\delta}(\tilde{\mathbf p})&=C(d)\,\left[
\delta_{\alpha\beta,\gamma\delta}^{\mathrm{Kr}}-
\Theta\big(\langle\langle
S^{\alpha}S^{\beta}S^{\gamma}S^{\delta}\rangle\rangle -
\langle\langle S^{\alpha}S^{\beta}\rangle\rangle
\langle\langle S^{\gamma}S^{\delta}\rangle\rangle\big)\right]\notag\\[3pt]
\label{mass}
&+\tilde{\mathbf p}^2\, \delta_{\alpha\beta,\gamma\delta}^{\mathrm{Kr}},
\end{align}
where $C(d)=2d\int f(r)^2\,d^dr/\int r^2f(r)^2\,d^dr$ is a smooth
function of dimensionality, but independent of the temperature and
magnetic field. As such, its concrete value is irrelevant, and a simple
adjustment of the cutoff function $f(r)$ can make it equal to unity.

A replica symmetric field-theory---for the study of the massive
high-temperature phase, and/or the massless critical manifolds---can be
obtained by choosing a replica symmetric mean-field value $Q_{\alpha\beta}
\equiv Q$ in Eq.~(\ref{Qsep}). The stationarity condition $\mathcal{L}
^{(1)}\equiv 0$ gives us an implicit equation for $Q$ (see Eq.~(\ref{L1})):
\begin{equation}\label{Q}
\Theta^{-1}Q=\langle\langle S^{\alpha}S^{\beta}\rangle\rangle =
\frac{\underset{\{S^{\alpha}\}}{\mathrm {Tr}}\left(S^{\alpha}S^{\beta}
e^{\sum_{(\alpha\beta)}Q\,S^{\alpha}S^{\beta}+
\frac{H}{kT}\sum_\alpha S^{\alpha}}\right)}
{\underset{\{S^{\alpha}\}}{\mathrm {Tr}}\left(
e^{\sum_{(\alpha\beta)}Q\,S^{\alpha}S^{\beta}+
\frac{H}{kT}\sum_\alpha S^{\alpha}}\right)}
\:\:,\quad \alpha\not=\beta.
\end{equation}
$Q$ enters the definition of the mass operator, Eq.~(\ref{mass}),
and the cubic interaction operator, Eq.~(\ref{W}), through the
Boltzmann-weight in the averages $\langle\langle\ldots\rangle\rangle$.
Replica symmetry is induced also for these operators, resulting in the
three different components of the mass:
\begin{align}
M_{\alpha\beta,\alpha\beta}(\tilde{\mathbf p})&=M_1+\tilde{\mathbf p}^2,
\notag\\
M_{\alpha\gamma,\beta\gamma}(\tilde{\mathbf p})&=M_2\,,\label{RSmass}\\
M_{\alpha\beta,\gamma\delta}(\tilde{\mathbf p})&=M_3\,;\notag
\end{align}
and the eight different components of the cubic interaction operator:
\begin{xalignat}{4}
W_{\alpha\beta,\beta\gamma,\gamma\alpha}&=W_1\, ,&
W_{\alpha\beta,\alpha\beta,\alpha\beta}&=W_2\, ,&
W_{\alpha\beta,\alpha\beta,\alpha\gamma}&=W_3\, ,&
W_{\alpha\beta,\alpha\beta,\gamma\delta}&=W_4\, ,\notag\\\label{RSW}
W_{\alpha\beta,\alpha\gamma,\beta\delta}&=W_5\, ,&
W_{\alpha\beta,\alpha\gamma,\alpha\delta}&=W_6\, ,&
W_{\alpha\gamma,\beta\gamma,\delta\mu}&=W_7\, ,&
W_{\alpha\beta,\gamma\delta,\mu\nu}&=W_8\, .
\end{xalignat}

$\mathcal{L}^{(2)}$ of Eq.~(\ref{L2M}), together with Eq.~(\ref{RSmass}),
is equivalent with that of Eq.~(\ref{L2}), provided the two sets of masses
are related by the following expressions:
\begin{align}
\label{m1} m_1&=\frac{1}{2}(M_1-2M_2+M_3)\, ,\\
\label{m2} m_2&=M_2-M_3\, ,\\
\label{m3} m_3&=\frac{1}{4}M_3\, .
\end{align}
Similarly, Eqs.~(\ref{L3}) and (\ref{L3W}) are two different
representations of $\mathcal{L}^{(3)}$. Using Eq.~(\ref{RSW}), a
one to one correspondence between the two sets of cubic couplings,
$w$'s and $W$'s, can be deduced by a somewhat lengthy but elementary
calculation. The results are listed in Table~1.
\begin{table}
\begin{align*}
w_1&=W_1-3W_5+3W_7-W_8\\
w_2&=\frac{1}{2}W_2-3W_3+\frac{3}{2}W_4+3W_5
+2W_6-6W_7+2W_8\\
w_3&=3W_3-3W_4-6W_5-3W_6+15W_7-6W_8\\
w_4&=\frac{3}{4}W_4-\frac{3}{2}W_7
+\frac{3}{4}W_8\\
w_5&=3W_5-6W_7+3W_8\\
w_6&=W_6-3W_7+2W_8\\
w_7&=\frac{3}{2}W_7-\frac{3}{2}W_8\\
w_8&=\frac{1}{8}W_8
\end{align*}
\caption{The relationship between the cubic couplings $w$'s of
Eq.~(\ref{L3}) and $W$'s of Eq.~(\ref{L3W}).}
\end{table}

\section{Analysis of the stationarity conditions and bare masses}
\label{analysis}

It is easy to recognize that, after a simple rescaling  of the temperature,
Eq.~(\ref{Q}) coincides with the replica symmetric mean-field equation
of Sherrington and Kirkpatrick (SK) \cite{SK,KS} for the order parameter of the
Ising spin glass on a fully-connected lattice. This may not be surprising:
the most direct way to define mean-field theory on a d-dimensional
hypercubic lattice is letting $\rho$, the range of interaction,
go to infinity, thus neglecting all the loop corrections to the equation
of state \cite{Amit,Brezinetal}. The solution $Q$ of Eq.~(\ref{Q}) has,
however, an application that goes beyond mean-field theory: it enters
the mass operator and interaction components in the formulae
Eqs.~(\ref{mass}) and (\ref{W}), respectively, through the effective average
$\langle\langle\ldots\rangle\rangle$. Although a field-theory emerging from
this procedure has a direct connection to the original parameters, such
as temperature, magnetic field and also replica number $n$, renormalisation
will reshuffle the masses and couplings,
possibly forcing them into some fixed point.

Following Refs.~\cite{SK,KS}, Eq.~(\ref{Q}) can be cast into a more
convenient form: 
\begin{align}\label{Qvar}
\Theta^{-1}Q&=\frac{\int\frac{du}{\sqrt{2\pi}}e^{-\frac{u^2}{2}}
\tanh^2(\sqrt{Q}u+H/kT)\cosh^{n}(\sqrt{Q}u+H/kT)}
{\int\frac{du}{\sqrt{2\pi}}e^{-\frac{u^2}{2}}\cosh^{n}(\sqrt{Q}u+H/kT)}\\[6pt]
&= \overline{\tanh^2(\sqrt{Q}u+H/kT)}.\notag
\end{align}
The shorthand notation
\begin{equation}\label{tanhaverage}
\overline{\tanh^k(\ldots)}\equiv\frac{\int\frac{du}{\sqrt{2\pi}}e^{-\frac{u^2}{2}}
\tanh^k(\ldots)\cosh^{n}(\sqrt{Q}u+H/kT)}
{\int\frac{du}{\sqrt{2\pi}}e^{-\frac{u^2}{2}}\cosh^{n}(\sqrt{Q}u+H/kT)}
\end{equation}
has been introduced for later use. By Eq.~(\ref{Qvar}), $Q$ is implicitly
given as a function of temperature, magnetic field and replica number.
One can easily find the SK spin glass transition point
($\Theta=1$, $H=0$) as an isolated singularity for {\em any\/} given $n$
close to
zero. (Keeping $n$ finite is for later use. At the moment, we must notice
that this singularity is rather unaffected by the $n\to 0$ limit.)
The relevant, positive, solution for $H=0$ is
\begin{equation}\label{SKQ}
Q=
\begin{cases}0& \text{for $t>0$},\\
             -\frac{1}{2-n}t
+\left[-\frac{n-3}{(n-2)^2}+\frac{1}{3(n-2)^3}\right]t^2
+\ldots& \text{for $t<0$};
\end{cases}
\end{equation}
where the new temperature scale $t\equiv \Theta^{-1}-1>0$ ($<0$) in the
disordered (spin glass) phase, respectively. As displayed in Figure~1,
one can join up smoothly the two regimes by-passing the critical point, like
in ordinary critical phenomena.\footnote{
It is obvious from Eq.~(\ref{Qvar}) that $Q=0$ is always a solution for zero
magnetic field, independently of $t$ and $n$. The $Q=0$ solution for $t<0$,
however, defines the negative $Q$ branch (starting from point G
in Figure~1(a) with $Q=0$, and following the dotted curve, you end up at P with
$Q<0$) which is
non-physical. The two branches meet at the SK transition point.}
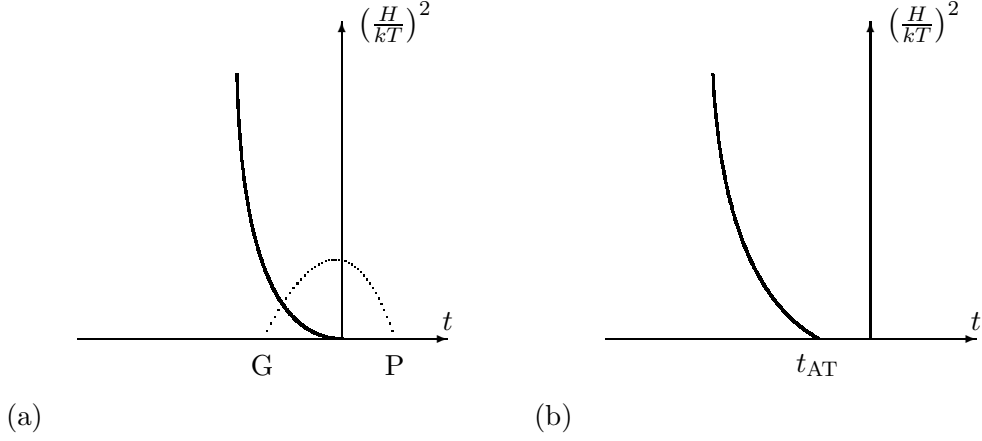
\begin{figure}\begin{center}
\begin{picture}(400,200)(-200,-100)
\put(140,-60){\vector(0,1){120}}
\put(40,-60){\vector(1,0){140}}
\put(160,60){\makebox(0,0){$\left(\frac{H}{kT}\right)^2$}}
\put(180,-50){\makebox(0,0)[t]{$t$}}
\put(-60,-60){\vector(0,1){120}}
\put(-160,-60){\vector(1,0){140}}
\put(-40,60){\makebox(0,0){$\left(\frac{H}{kT}\right)^2$}}
\put(-20,-50){\makebox(0,0)[t]{$t$}}\put(120,-70){\makebox(0,0){$t_{\text{AT}}$}}
\put(20,-90){\makebox(0,0){(b)}}\put(-180,-90){\makebox(0,0){(a)}}
\qbezier[40](-40,-60)(-60,0)(-90,-60)\put(-40,-70){\makebox(0,0){P}}
\put(-90,-70){\makebox(0,0){G}}
\thicklines \qbezier(-60,-60)(-98,-60)(-100,40)
\qbezier(120,-60)(84,-40)(80,40)
\end{picture}
\end{center}\caption{(a): Mean-field phase diagram for $n=0$. Starting from the
paramagnetic state P with $Q=0$, and following the dotted curve, the SK critical
point ($t=H=0$) can be by-passed ending in the glassy state G with $Q>0$.
(b): Mean-field phase diagram for $n\gtrsim 0$. There is a temperature range even
in zero field,
$t_{\text{AT}}<t<0$, where $Q$ is positive and the replica symmetric state is stable.}
\end{figure}

To go beyond a mean-field solution, and build loops, one must have
well-defined Gaussian propagators,
i.e.\ the eigenvalues of the mass operator must be
non-negative. A generic replica symmetric mass operator, like that in
Eq.~(\ref{RSmass}), was diagonalised years ago \cite{PR,BrMo} with the
following expressions for the eigenvalues (which will play the role of bare
masses here):
\begin{align}
\label{rR} r_R&=M_1-2M_2+M_3,\\
\label{rA} r_A&=M_1+(n-4)M_2-(n-3)M_3,\\
\label{rL} r_L&=M_1+2(n-2)M_2+\frac{(n-2)(n-3)}{2}M_3.
\end{align}
The indices $R$, $A$ and $L$ stand for replicon, anomalous
and longitudinal, respectively, each referring to the corresponding
family of eigenmodes (see Ref.\ \cite{BrMo} and also later sections). From Eqs.~(\ref{mass})
and (\ref{RSmass}), and using the stationarity condition (\ref{Q}),
the following expressions are  obtained for the bare masses:
\begin{align}
\label{rRvar} r_R&=\frac{t}{1+t}+2Q-\frac{1}{1+t} 
\, \overline{\tanh^4(\sqrt{Q}u+H/kT)},\displaybreak[0]\\[6pt]
\label{rAvar} r_A&=r_R+(n-2)\left[\frac{1}{1+t}
\,\overline{\tanh^4(\sqrt{Q}u+H/kT)}-Q\right],\displaybreak[0]\\[6pt]
\label{rLvar} r_L&=r_R+\frac{n-1}{2}\left[-(n-4)\frac{1}{1+t}
\,\overline{\tanh^4(\sqrt{Q}u+H/kT)}-4Q+n(1+t)Q^2\right].
\end{align}
An expansion below the transition provides, in zero field,
\begin{align}
\label{rRexp} r_R&=\frac{n}{2}(1+\frac{n}{2}+\ldots)\,(-t)
+(-\frac{1}{3}+\dots)\,t^2+\ldots,\\
\label{rAexp} r_A&=(1+\frac{n}{2}+\ldots)\,(-t)+(-\frac{5}{12}+\dots)\,t^2
+\ldots,\\
\label{rLexp} r_L&=(-t)+(-\frac{5}{12}+\dots)\,t^2+\ldots.
\end{align}
(The $n$-dependent coefficients are displayed in an expanded form
for showing clearly the signs for small $n$. The complete $n$-dependence,
however, is easily found.) The hitherto degenerate masses split
when passing the SK transition singularity (where they are zero) and
emerge positively in the spin glass phase for any $n\gtrsim 0$. $r_R$
starts, however, with a small slope proportional to $n$, and becomes
massless again at the AT surface
$t_{\text{AT}}=-\frac{3}{2}n+\ldots$, where instability of the replica
symmetric phase begins. This result was first derived by Kondor
\cite{Kondor}, the aspect we wish to emphasize here is the existence of
an intermediary temperature range where replica symmetry persists,
see Figure~1,
though, as a result of a nonzero $Q$, the level of symmetry is lower,
leading, at the mean-field level, to the splitting of the bare masses.

By definition, $r_R\equiv 0$ on the AT-surface. From Eqs.~(\ref{Qvar})
and (\ref{rRvar}), the magnetic field can be expressed as a double series
for small $-t$ and $n$; the leading, cubic, term is as follows:
\[(H/kT)^2=\frac{1}{6}(-t)^3-\frac{1}{4}nt^2=-\frac{3}{8}n^2(t-t_{\text{AT}})
+\frac{1}{2}n(t-t_{\text{AT}})^2-\frac{1}{6}(t-t_{\text{AT}})^3.\]
This can be cast into a scaling form
\begin{equation}\label{ATscaling}
(H/kT)^2=t^{\tau}\varphi(\frac{n}{t^{\kappa}}),\qquad t,n\to 0,
\end{equation}
where the exponents have now their mean-field values $\tau=3$
and $\kappa=1$, and $\varphi(\dots)$ is the scaling function
characterising the AT-surface.
 
On the basis of the above mean-field analysis, we can define two
different, though both replica symmetric, cubic field-theories relevant
to describe spin glass transitions of different types in low enough
dimensions ($d$ certainly smaller than eight):
\begin{itemize}
\item
The zero-field model---$H\equiv 0,t\geq 0$ and $n\gtrsim 0$---implies,
through Eq.~(\ref{SKQ}), $Q\equiv 0$, leading to degenerate bare masses
$r_R=r_A=r_L$, see Eqs.~(\ref{rRvar}), (\ref{rAvar}) and (\ref{rLvar}).
It is easy to verify using Eqs.~(\ref{W}) and (\ref{RSW}) that all the
$W$'s but $W_1$ are zero, and the same is true, by Table 1, for the small
$w$'s. Let us put down explicitly the definition of the cubic field
theory for the zero-field transition from the paramagnetic
phase:
\begin{equation}\label{H=0model}
\begin{aligned}
r\equiv& r_R=r_A=r_L;\\
w\equiv& w_1,\qquad w_i=0,\quad i=2,\ldots,8.
\end{aligned}
\end{equation}
At the mean-field level, a massless state is reached along the critical
{\em line\/} $H=0$, $t=0$ and $n\gtrsim 0$; the replica number being a rather
innocent parameter around zero. We expect this property to remain
for short-ranged (finite-dimensional) systems, and indeed, $\epsilon
$-expansion results have supported this idea \cite{epsilon}.    
(This kind of replica field-theory was studied in
all the Refs.~\cite{Harris,CheLu,Elderfield,PR,BrMo,epsilon}.)
\item
The second model, which is in fact the most general cubic field-theory
with an unbroken replica symmetry, has all the three masses $m_i$ and
eight couplings $w_i$ different. At the mean-field level, it corresponds
to a nonzero $Q$ which is always such when a magnetic field is
switched on. More surprisingly, however, there is a whole range of
temperatures $t_{\text{AT}}\leq t < 0$ even in zero field where the
bare parameters correspond to this more general model. Criticality
is induced, at least at the mean-field level, by the masslessness of
the replicon mode on the de Almeida-Thouless {\em surface\/}.
The replica number $n$
is now a crucially important parameter; a fact clearly shown by
the scaling formula (\ref{ATscaling}). How fluctuations will modify
this picture is a prime problem in spin glass theory. There has been
two attempts, at least to our knowledge, adressing this question
\cite{BrRo,Greenetal}. In Ref.~\cite{BrRo} fluctuations were
restricted to the replicon subspace; in our language this means for the masses
that
$r_A=r_L=\infty$ and $r_R$ critical, while all the cubic couplings were zero
but $w_1$ and $w_2$. The effect of a small magnetic field was
introduced in Ref.~\cite{Greenetal} by shifting the bare masses from
their zero-field values (more precisely, beside $m_1$, $m_2$ became massive
too),
the couplings remained, however, unchanged.
The replica number was effectively set to zero in these works.
Here we wish to emphasize the role $n$ may play in a search for a
de Almeida-Thouless transition
in finite-dimensional systems.
\end{itemize}

\section{A canonical representation of the cubic interaction}
\label{canonical}

In a field-theory with more than one mass, as in the generic replica
symmetric system introduced in the previous sections, critical manifolds
can be classified by their massless eigenmodes. In this case, it is more
convenient to use directly the {\em eigenvalues\/} of the mass operator ($r_R$,
$r_A$ and $r_L$; Eqs.~(\ref{rR}), (\ref{rA}) and (\ref{rL})), instead of
the sets $m$'s or $M$'s. At that point it is natural to ask what will
happen with the interaction vertices whose legs join the, by this time
block-diagonalized, propagators.
We show in this section how the transformation that block-diagonalizes
the quadratic part of the Lagrangean into ``modes'' induces a new set of
cubic couplings describing how these modes interact.
Our replicated field-theory
becomes more tractable after using these ``canonical'' cubic
parameters.

The  $\frac{1}{2}n(n-1)$-dimensional vectorspace
spanned by the two-replica fields $\phi^{\alpha\beta}$ has the simple
structure being a direct sum of the subspaces called longitudinal,
anomalous and replicon. Their definitions are as follows ($\phi^{\alpha\beta}
=\phi^{\beta\alpha}$ and $\phi^{\alpha\alpha}=0$ are understood everywhere,
of course):
\begin{itemize}
\item The longitudinal (L) subspace consists of replica symmetric vectors,
i.e.\ independent of replica indices. Each element from this subspace
corresponds to a {\em scalar} $\phi\,$:
\begin{equation}\label{Lsubspace}
\phi_{L}^{\alpha\beta}=\phi,
\end{equation}
and it is obviously one-dimensional.
\item Any element of the anomalous (A) subspace can be represented by
a one-replica field $\phi^{\alpha}$, i.e.\ by a {\em vector\/} restricted,
however, by the condition
\begin{equation}\label{Acondition}
\sum_{\alpha}\phi^{\alpha}=0.
\end{equation}
A generic anomalous field can now be written as
\begin{equation}\label{Asubspace}
\phi_{A}^{\alpha\beta}=\frac{1}{2}(\phi^{\alpha}+\phi^{\beta}).
\end{equation}
As a result of condition (\ref{Acondition}), the anomalous subspace is
$n-1$-dimen\-sion\-al.
\item True two-replica fields, loosely speaking {\em tensors}, constitute
the replicon (R) subspace with the restriction
\begin{equation}\label{Rcondition}
\sum_{\beta}\phi_{R}^{\alpha\beta}=0\quad \text{for any}\quad \alpha
=1\ldots n.
\end{equation}
From the $n$ equations above follows that the number of independent
$\phi_{R}^{\alpha\beta}$ is $\frac{1}{2}n(n-1)-n=\frac{1}{2}n
(n-3)$, rendering the replicon subspace $\frac{1}{2}n(n-3)$-dimensional.
\end{itemize}
A generic field $\phi^{\alpha\beta}$ can always be decomposed into the
sum
\begin{equation}\label{dsum}
  \phi^{\alpha\beta}=\phi_{L}^{\alpha\beta}+
                     \phi_{A}^{\alpha\beta}+
                     \phi_{R}^{\alpha\beta}.
\end{equation}

It is straightforward to see that the subspaces defined above give the
exact diagonalisation of a {\em generic} replica symmetric matrix, as
defined in the equations of (\ref{RSmass}), i.e.
\begin{equation}\label{eigenvalueeq}
         \sum_{(\gamma\delta)} M_{\alpha\beta,\gamma\delta}
           \phi_{i}^{\gamma\delta}=r_{i}
           \phi_{i}^{\alpha\beta}\,,\qquad i=L,A,R\:\,,
\end{equation}  
the eigenvalues given in Eqs.~(\ref{rL}), (\ref{rA}) and (\ref{rR}),
respectively.
For a generalization to higher order operators, the matrix-element
of $M$ between two arbitrary vectors $\phi^{\alpha\beta}$ and $\psi
^{\alpha\beta}$ will be computed after having represented them by
the longitudinal scalar, anomalous  vector and
replicon tensor, as explained above (see Eqs.~(\ref
{Lsubspace}), (\ref{Asubspace}) and (\ref{dsum})). An expression in
terms of the three second-order invariants $\phi \psi$,
$\sum_{\alpha}\phi^{\alpha}\psi^{\alpha}$ and $\sum_{\alpha\beta}
\phi^{\alpha\beta}_{R}\psi^{\alpha\beta}_{R}$ arises:
\begin{equation}\label{matrixelement}
\begin{split}
M_{\{\phi,\psi\}}&\equiv
\sum_{(\alpha\beta),(\gamma\delta)} M_{\alpha\beta,\gamma\delta}
\phi^{\alpha\beta}\psi^{\gamma\delta}\\
&=\frac{1}{2}r_R\sum_{\alpha\beta}
\phi^{\alpha\beta}_{R}\psi^{\alpha\beta}_{R}+
\frac{n-2}{4}r_A\sum_{\alpha}\phi^{\alpha}\psi^{\alpha}+
\frac{n(n-1)}{2}r_L\phi \psi
\end{split}
\end{equation}
(to get the anomalous term, the restriction (\ref{Acondition}) has
been used). Except for numerical factors, the corresponding eigenvalues
appear as the coefficients of the three possible second order invariants;
viz.\ RR, AA and LL. 

The most important point we can learn from Eq.~(\ref{matrixelement}) is
a complete factorization of a matrixelement, provided $\phi$ and $\psi$
are chosen from one of the subspaces L, A or R; i.e.
\begin{multline}\label{factorization}
M_{\{\phi,\psi\}}=\\
\{\text{expression of mass components}\}
\times \{\text{invariant composed of $\phi$ and $\psi$}\}.
\end{multline}
Generalization to a cubic replica symmetric operator, as defined
in Eq.~(\ref{RSW}), is straightforward. The analogue of the matrixelement
can be easily defined by
\begin{equation}\label{cubicvertex}
W_{\{\phi,\psi,\chi\}}\equiv
\sum_{(\alpha\beta),(\gamma\delta),(\mu\nu)}
W_{\alpha\beta,\gamma\delta,\mu\nu}\phi^{\alpha\beta}\psi^{\gamma\delta}
\chi^{\mu\nu}.
\end{equation}
Taking each of the fields $\phi$, $\psi$ and $\chi
$ from one of the subspaces L, A or R, the nonzero values 
obtained can be listed
as follows:
\begin{equation}\label{canonicalform}
\begin{aligned}
W_{RRR}&=g_1\,\sum_{\alpha\beta\gamma}\phi^{\alpha\beta}\psi^{\beta\gamma}
\chi^{\gamma\alpha}+\frac{1}{2}g_2\,\sum_{\alpha\beta}
\phi^{\alpha\beta}\psi^{\alpha\beta}\chi^{\alpha\beta},\\
W_{RRA}&=g_3\,\sum_{\alpha\beta}
\phi^{\alpha\beta}\psi^{\alpha\beta}\chi^{\alpha},\\
W_{RRL}&=g_4\,\sum_{\alpha\beta}
\phi^{\alpha\beta}\psi^{\alpha\beta}\chi,\\
W_{RAA}&=g_5\,\sum_{\alpha\beta}
\phi^{\alpha\beta}\psi^{\alpha}\chi^{\beta},\\
W_{AAA}&=g_6\,\sum_{\alpha}\phi^{\alpha}\psi^{\alpha}\chi^{\alpha},\\
W_{AAL}&=g_7\,\sum_{\alpha}\phi^{\alpha}\psi^{\alpha}\chi,\\
W_{LLL}&=g_8\,\phi\psi\chi.
\end{aligned}
\end{equation}
Symmetry makes the remaining $W_{RAL}$, $W_{RLL}$ and $W_{ALL}$ all
identically zero. As for the masses, a complete factorization occurs in
the above formula, except the RRR vertex.\footnote{The reason for that
is the two different cubic invariants we can construct from
replicon fields; see the first line of Eq.~(\ref{canonicalform}).}
We prefer the name ``canonical'' for the set of 
cubic parameters $g$ emerging in
the above formulas, as they are, in some sense, an extension of the
notion of eigenvalues to the cubic interaction term. After a somewhat
lengthy calculation, we obtained the set of equations
for the $g_i$ in terms of the $W_i$, $i=1,\ldots,8$,
which are the
counterparts of Eqs.~(\ref{rR}), (\ref{rA}) and (\ref{rL}).
(We omit to display these rather complicated, and not very
instructive, expressions here; they can be easily obtained from
Eqs.\ (49a-h) below and using Table 1.)

Comparing Eqs.~(\ref{L3}) and (\ref{canonicalform}), a one to one
correspondence between a $w_i$ and a $g_i$, $i=1,\ldots,8$, is obvious.
When expressing the $g$'s in terms of the $w$'s, instead of the $W$'s,
not only the formulas become simpler, but a clear R $\rightarrow$
A $\rightarrow$ L hierarchy emerges:\addtocounter{equation}{1}
\begin{align}
g_1&=w_1,\tag{\theequation a}\displaybreak[0]\\
g_2&=2w_2,\tag{\theequation b}\displaybreak[0]\\
g_3&=-w_1+w_2+\frac{n-2}{6}w_3,\tag{\theequation c}\displaybreak[0]\\
g_4&=-w_1+w_2+\frac{n-1}{3}w_3+\frac{n(n-1)}{3}w_4,\tag{\theequation d}\displaybreak[0]\\
g_5&=\frac{n-4}{4}w_1+\frac{1}{2}w_2+\frac{n-2}{6}w_3
+\frac{(n-2)^2}{12}w_5,\tag{\theequation e}\displaybreak[0]\\
g_6&=-\frac{3n-8}{4}w_1+\frac{n-4}{4}w_2+\frac{(n-2)(n-4)}{8}w_3
-\frac{(n-2)^2}{4}w_5+\frac{(n-2)^3}{8}w_6,\tag{\theequation f}\displaybreak[0]\\
g_7&=\frac{n-2}{2}\bigg[
\frac{n-4}{2}w_1+w_2+\frac{2n-3}{3}w_3
+\frac{n(n-1)}{3}w_4\notag\\&+\frac{(n-2)(2n-3)}{6}w_5
+\frac{n-1)(n-2)}{2}w_6+\frac{n(n-1)(n-2)}{6}w_7\bigg],\tag{\theequation g}\displaybreak[0]\\
g_8&=n(n-1)\Big[
(n-2)w_1+w_2+(n-1)w_3+n(n-1)w_4\notag\\&+(n-1)^2w_5+(n-1)^2w_6
+n(n-1)^2w_7+n^2(n-1)^2w_8\Big].\tag{\theequation h}
\end{align}

\section{Illustration of the technique: one-loop calculation of the
masses}\label{illustration}

We apply the standard definition of the mass operator, i.e.\ it is
the zero momentum limit of the inverse of the two-point function:
\begin{align}\label{Gamma}
\Gamma_{\alpha\beta,\gamma\delta}&\equiv\lim_{\mathbf p\to 0}
\big[G^{-1}(\mathbf p)\big]_{\alpha\beta,\gamma\delta}\\
\intertext{where the {\em connected\/} two-point function $G$ is the average}
\big[G(\mathbf p)\big]_{\alpha\beta,\gamma\delta}&\equiv\label{two-point}
\langle \phi^{\alpha\beta}_{\mathbf p}\phi^{\gamma\delta}_{-\mathbf p}
\rangle
-\langle \phi^{\alpha\beta}_{\mathbf p}\rangle
\langle\phi^{\gamma\delta}_{-\mathbf p}\rangle
\end{align}
taken with the statistical weight $\sim e^{-\cal L}$, ${\cal L}={\cal L}^{(2)}
+{\cal L}^{(3)}$; see Eqs.~(\ref{L2}) and (\ref{L3}).
Dyson's equation for $\Gamma$ allows us to compute it perturbatively:
\begin{equation}\label{Dyson}
\Gamma_{\alpha\beta,\gamma\delta}=M_{\alpha\beta,\gamma\delta}(\tilde
{\mathbf p}=0)-\Sigma_{\alpha\beta,\gamma\delta}(\tilde
{\mathbf p}=0),
\end{equation}
where the mean-field mass operator $M_{\alpha\beta,\gamma\delta}$ has
been defined in Eqs.\  (\ref{mass}) and (\ref{RSmass}), whereas the
self-energy $\Sigma$ contains all the one-particle irreducible
graphs to the two-point function, with external lines omitted.
Up to leading, one-loop, order it is given as the simple ``bubble''
diagram:
\begin{multline}
\Sigma_{\alpha\beta,\gamma\delta}(\tilde
{\mathbf p}=0)=\\[10pt]
\label{bubble}
\frac{1}{2z}\int^{\Lambda}\frac{d^d \tilde p}{(2\pi)^d}
\sum
W_{\alpha\beta,\alpha'\beta',\alpha''\beta''}
G_{\alpha'\beta',\gamma'\delta'}^{(0)}(\tilde p)
G_{\alpha''\beta'',\gamma''\delta''}^{(0)}(\tilde p)
W_{\gamma\delta,\gamma'\delta',\gamma''\delta''}
\end{multline}
where the free propagator $G^{(0)}$ is defined by
\begin{equation}\label{freepropagator}
G_{\alpha\beta,\gamma\delta}^{(0)}(\tilde p)\equiv
\big[M^{-1}(\tilde p)\big]_{\alpha\beta,\gamma\delta}\,,
\end{equation}
see Eqs.\  (\ref{mass}) and (\ref{RSmass}),
and the bare vertices, $W$'s, have been introduced in Eqs.\  (\ref{L3W}),
(\ref{W}) and (\ref{RSW}).

To compute the replica sum in Eq.\  (\ref{bubble}), we must overcome the
problem of having non-diagonal free propagators. This project can be
easily accomplished by calculating $\Sigma_{m,n}$, instead of
$\Sigma_{\alpha\beta,\gamma\delta}$, defined as:
\begin{equation}\label{mn}
\Sigma_{m,n}\equiv \sum_{(\alpha\beta),(\gamma\delta)}\phi_m^{\alpha\beta}
\,\Sigma_{\alpha\beta,\gamma\delta}\,\phi_n^{\gamma\delta},
\end{equation}
$\phi_m$ and $\phi_n$ taken from the non-orthogonal basis discussed in
\appendixname\ A. Exploiting the completeness of both this basis and its
biorthogonal counterpart\footnote{What we use in the derivation of
this formula is the decomposition of the unit operator:
$\delta^{\text{Kr}}_{\alpha\beta,\gamma\delta}=\sum_m \phi_m^{\alpha\beta}
{\tilde\phi}_m^{\gamma\delta}=\sum_m{\tilde\phi}_m^{\alpha\beta}
\phi_m^{\gamma\delta}$.}, we can transform
$\Sigma_{m,n}$ into a representation with {\em diagonal} free propagators:
\begin{equation}\label{diagonal}
G^{(0)}_{m,\tilde n}\equiv \sum_{(\alpha\beta),(\gamma\delta)}
\phi_m^{\alpha\beta}\,G^{(0)}_{\alpha\beta,\gamma\delta}
\,{\tilde\phi}_n^{\gamma\delta}=
G^{(0)}_{m}\,\delta^{\text{Kr}}_{mn}
\end{equation}
(a ``tilde'' refers always to a member of the reciprocal basis).
$G^{(0)}_{m}$, as it is the eigenvalue of the free propagator matrix,
can be simply related, through Eq.\ (\ref{freepropagator}), to one of the
three bare masses of Eqs.\ (\ref{rR}), (\ref{rA}) and (\ref{rL}):
\begin{equation}\label{G0}
G^{(0)}_{m}(\tilde p)=\frac{1}{r_m+{\tilde p}^2},
\end{equation}
$r_m=r_R$, $r_A$ or $r_L$ depending on the subspace $m$ belongs to (see
\appendixname\ A).

We can now propose a simple graphical representation for $\Sigma_{m,n}$
by introducing arrowed lines for the free propagators $G^{(0)}$
joining interaction vertices $W$:
\[
\begin{gathered}G^{(0)}_{m}\quad\Leftrightarrow\quad\end{gathered}
\begin{gathered}
\begin{picture}(80,0)\put(0,0){\vector(1,0){80}}\put(40,5){\makebox(0,0){$m$}}
\end{picture}\end{gathered}\qquad.
\]
Using the convention of \appendixname\ B concerning the meaning of inward
and outward arrows, we can draw for $\Sigma_{m,n}$:
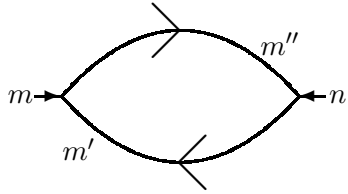
\[\begin{picture}(200,100)(-100,-50)
{\thicklines\put(0,25){\line(-1,1){10}}\put(0,25){\line(-1,-1){10}}
\put(0,-25){\line(1,1){10}}\put(0,-25){\line(1,-1){10}}
\qbezier(-45,0)(0,50)(45,0)
\qbezier(-45,0)(0,-50)(45,0)
\put(-55,0){\vector(1,0){10}}\put(55,0){\vector(-1,0){10}}}
\put(-60,0){\makebox(0,0){$m$}}
\put(60,0){\makebox(0,0){$n$}}\put(-38,-20){\makebox(0,0){$m'$}}
\put(38,20){\makebox(0,0){$m''$}}
\end{picture}
\]
which can be spelled out explicitly as:
\begin{equation}\label{Sigma}
\Sigma_{m,n}=
\frac{1}{2z}\int^{\Lambda}\frac{d^d \tilde p}{(2\pi)^d}
\sum_{m',m''}W_{m,m',{\tilde m}''}\,W_{n,m'',{\tilde m}'}\,
G^{(0)}_{m'}(\tilde p)\,G^{(0)}_{m''}(\tilde p).
\end{equation}
(Summations are over the set of $n(n-1)/2$ modes of
\appendixname\ A.) Orthogonality of the different subspaces restricts
the number of nonzero elements of the matrix $\Sigma_{m,n}$ to the
cases where $m$ and $n$ belong to the same family R, A or L. To compute
the three eigenvalues of the self-energy, we can make the simplest
possible choices for $m$ and $n$, i.e.
\begin{align}
\label{SigmaR}\Sigma_{(\mu\nu),(\mu\nu)}&=\Sigma_{R}\,
\sum_{(\alpha\beta)}\phi^{\alpha\beta}_{(\mu\nu)}\phi^{\alpha\beta}_{(\mu\nu)}=
\frac{(n-1)(n-2)^2(n-3)}{4}\:\Sigma_{R}\,,\\
\label{SigmaA}\Sigma_{(\mu),(\mu)}&=\Sigma_{A}\,
\sum_{(\alpha\beta)}\phi^{\alpha\beta}_{(\mu)}\phi^{\alpha\beta}_{(\mu)}=
\frac{n(n-1)(n-2)}{4}\:\Sigma_{A}\,,\\
\label{SigmaL}\Sigma_{(L),(L)}&=\Sigma_{L}\,
\sum_{(\alpha\beta)}\phi^{\alpha\beta}_{(L)}\phi^{\alpha\beta}_{(L)}=\frac{n(n-1)}{2}\:\Sigma_{L}\,.
\end{align}
(The computation of the scalar products above is relatively easy using the
definitions of the basis functions in \appendixname\ A.) An extensive
use of the table of cubic vertices
$W_{m,m',{\tilde m}''}$ in \appendixname\ B makes it possible
to compute the left-hand sides; the feasibility of the calculation is,
however, due to the {\em selection rule} we explain in that appendix.
The results can be summarized by displaying the eigenmodes of the mass
operator $\Gamma$, by means of Eq.~(\ref{Dyson}), valid to first order
in $1/z$:
\begin{align}
\Gamma_{R}&=r_R
-\bigg\{\Big[\frac{n^4-8n^3+19n^2-4n-16}{(n-1)(n-2)^2}
\,g_1^2+\frac{2(3n^2-15n+16)}{(n-1)(n-2)^2}\,g_1g_2\notag\\
&+\frac{n^3-9n^2+26n-22}
{2(n-1)(n-2)^2}\,g_2^2\Big]\,I_{RR}\notag\\
&+\frac{8(n-1)(n-4)}{n(n-2)^2}\,
g_3^2\,I_{RA}+\frac{8}{n(n-1)}\,g_4^2\,I_{RL}+\frac{16}
{(n-2)^2}\,g_5^2\,I_{AA}\bigg\}\label{GammaR}\,;\displaybreak[0]\\
\Gamma_{A}&=r_A-\bigg\{\frac{2(n-3)(n-4)}{(n-2)^2}\,g_3^2\,
I_{RR}+\frac{16n(n-3)}{(n-1)(n-2)^2}\,g_5^2\,I_{RA}\notag\\
&+\frac{32}{n(n-2)^2}\,g_6^2\,I_{AA}+\frac{32}{n(n-1)(n-2)^2}
\,g_7^2\,I_{AL}\bigg\}\label{GammaA}\,;\displaybreak[0]\\
\Gamma_{L}&=r_L-\bigg\{\frac{2(n-3)}{n-1}\,g_4^2\,I_{RR}
+\frac{16}{n(n-2)^2}\,g_7^2\,I_{AA}+\frac{4}{n^3(n-1)^3}
\,g_8^2\,I_{LL}\bigg\}\label{GammaL}\,.
\end{align}
To help the reader to understand the structure of the corrections to
the masses, we introduced the short-cut notation
\[
I_{ss'}\equiv\frac{1}{z}\int^{\Lambda}\frac{d^d \tilde p}{(2\pi)^d}
\,\frac{1}{r_s+{\tilde p}^2}\,\frac{1}{r_{s'}+{\tilde p}^2}
\]
for the momentum integrals; $s$ and $s'$ correspond to one of the subspaces
R, A or L. By means of the transformation rules between the $g$ and $w$ couplings,
Eqs.\ (49a-h), the above equations for the masses can be easily expressed in terms
of the $w$'s too.

\section*{Acknowledgements}

This work has been supported by the Hungarian Science Fund (OTKA),
Grant No.~T032424.

\appendix
\section*{\appendixname}
\section{A simple non-orthogonal basis}

For applying the canonical vertices defined in Eqs.~(\ref{cubicvertex})
and (\ref{canonicalform}), it is necessary to introduce a
basis in each of the subspaces; an obviously nonunique task. The
non-orthogonal basis defined below is not only the
simplest\footnote{This time, simplicity and orthogonality contradict each other.
An orthogonal, still rather complicated system was proposed years ago
\cite{Cwilich}.}, but cubic vertices evaluated in this basis will have a
remarkable property,
a kind of a selection rule involving replica numbers,
making computation of Feynmann-graphs feasible
(see \appendixname\ B). 

A member of this non-orthogonal basis will be denoted by $\phi^{\alpha
\beta}_{m}$, whereas its biorthogonal\footnote
{Biorthogonality has the usual
definition $\sum_{(\alpha\beta)}
\phi^{\alpha\beta}_{m}{\tilde\phi}^{\alpha\beta}_{m'}
=\delta^{\text{Kr}}_{mm'}$.}
counterpart, a member of the reciprocal basis, as  ${\tilde\phi}^{\alpha
\beta}_{m}$,
where $m$ \mbox{stands} for the modes in the subspaces L, A
and R as follows:
\begin{itemize}
\item The L subspace is one-dimensional, i.e.\ $m\leftrightarrow(L)$.
\item $m$ runs the single replica numbers, except one (which we
choose the $n^{\text{th}}$), for the $n-1$-dimen\-sional A space:
$m\leftrightarrow(\mu)$, $\mu=1,\ldots,n-1$.
\item In case of replicon modes,
$m$ corresponds to a pair of replicas, $m\leftrightarrow(\mu\nu)$,
with $\mu$,$\nu=1,\ldots,n-1$ and $\mu\not =\nu$. To ensure the correct
dimensionality $n(n-3)/2$ imposed by condition (\ref{Rcondition}), we
have to pick out an arbitrarily chosen pair $(\bar\mu\bar\nu)$,
giving for the number of replicon modes:
\[ \frac{(n-1)(n-2)}{2}-1=\frac{n(n-3)}{2}.\]
To sum up, there are two types of replicon modes:
\begin{equation}\label{Rmodes}
m\leftrightarrow
\begin{cases} (\mu\nu),&\mu,\nu=1,\ldots,n-1;\quad \mu,\nu\not=\bar\mu
\quad \text{or}\quad \bar\nu,\\
(\mu\bar\mu)\quad \text{or}\quad (\mu\bar\nu),&\mu=1,\ldots,n-1;\qquad
\quad \mu
\not=\bar\mu\quad \text{or}\quad \bar\nu.
\end{cases}
\end{equation}
\end{itemize}
In what follows we want to collect the results, omitting any proof.

\subsection*{L subspace:}

\begin{equation}\label{Lbasis}
\phi^{\alpha\beta}_{(L)}\equiv\phi_{(L)}=1;\qquad
{\tilde\phi}^{\alpha\beta}_{(L)}\equiv
{\tilde\phi}_{(L)}=\frac{2}{n(n-1)}.
\end{equation}

\subsection*{A subspace:}
\begin{equation}\label{Abasis}
\phi^{\alpha\beta}_{(\mu)}=
\begin{cases}1&\text{if $\alpha,\beta\not=\mu$},\\[5pt]
             -\frac{n-2}{2}&\text{if $\alpha=\mu$ or $\beta=\mu$};
\end{cases}\qquad
{\tilde\phi}^{\alpha\beta}_{(\mu)}=\frac{4}{n^2(n-2)}
\Big[\phi^{\alpha\beta}_{(\mu)}+
\sum_{\nu=1}^{n-1}\phi^{\alpha\beta}_{(\nu)}\Big].
\end{equation}
The one-replica objects, introduced in Eq.~(\ref{Asubspace}),
representing them are
\begin{equation}\label{onereplicabasis}
\phi^{\alpha}_{(\mu)}=
\begin{cases}1&\text{if $\alpha\not=\mu$},\\[5pt]
              -(n-1)&\text{if $\alpha=\mu$};
\end{cases}\qquad
{\tilde\phi}^{\alpha}_{(\mu)}=
\begin{cases}0&\text{if $\alpha\not=\mu,n$},\\[5pt]
             -\frac{4}{n(n-2)}&\text{if $\alpha=\mu$},\\[5pt]
              \frac{4}{n(n-2)}&\text{if $\alpha=n$}.
\end{cases}
\end{equation}
        
\subsection*{R subspace:}
\begin{equation}\label{Rbasis}
\phi^{\alpha\beta}_{(\mu\nu)}=
\begin{cases}1&\text{if $\alpha,\beta\not=\mu,\nu$}\,,\\[5pt]
             -\frac{n-3}{2}&\text{if there is one common replica
             index of the}\\
&\text{two pairs $\alpha,\beta$ and $\mu,\nu$}\,,\\[5pt]
             \frac{(n-2)(n-3)}{2}&\text{if the two pairs $\alpha,\beta$
             and $\mu,\nu$ are identical}.
\end{cases}
\end{equation}
In the reciprocal basis, we have different forms for the two types
of replicon modes defined in Eq.~(\ref{Rmodes}):
\begin{description}
\item[type-I]
\begin{equation}\label{type-I}
{\tilde\phi}^{\alpha\beta}_{(\mu\nu)}=
\begin{cases}\frac{2}{(n-1)(n-2)}&\text{if $(\alpha\beta)=
(\mu\nu)$ or $(\bar\mu n)$ or $(\bar\nu n)$},\\[5pt]
-\frac{2}{(n-1)(n-2)}&\text{if $(\alpha\beta)=
(\bar\mu\bar\nu)$ or $(\mu n)$ or $(\nu n)$},\\[5pt]
0&\text{otherwise};
\end{cases}
\end{equation}
\item[type-II]
\begin{equation}\label{type-II}
{\tilde\phi}^{\alpha\beta}_{(\bar\mu\mu)}=
\begin{cases}\frac{2}{(n-1)(n-2)}&\text{if $(\alpha\beta)=
(\bar\mu\mu)$ or $(\bar\nu n)$},\\[5pt]
-\frac{2}{(n-1)(n-2)}&\text{if $(\alpha\beta)=
(\bar\mu\bar\nu)$ or $(\mu n)$},\\[5pt]
0&\text{otherwise}.
\end{cases}
\end{equation}
\end{description}

\section{Vertex rules} 
For a cubic vertex with $\phi=\phi_m$, $\psi=\phi_{m'}$
and $\chi={\tilde\phi}_{m''}$ in Eq.~(\ref{cubicvertex}) ($m$, $m'$
and $m''$ referring to the modes introduced in \appendixname\ A),
a simple graphical representation can be given, namely
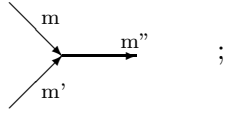
\[\begin{picture}(200,100)(-100,-50)
\put(-20,20){\vector(1,-1){20}}
\put(-20,-20){\vector(1,1){20}}
\put(0,0){\vector(1,0){28}}
\put(-8,12){\makebox(0,0)[bl]{\scriptsize m}}
\put(-8,-16){\makebox(0,0)[bl]{\scriptsize m'}}
\put(22,3){\makebox(0,0)[bl]{\scriptsize m''}}
\put(60,0){\makebox(0,0){;}}

\end{picture}\]
inward (outward) arrows correspond to ordinary (reciprocal) basis
functions, respectively. Such vertices have the remarkable property,
a kind of a {\em selection rule}, that the replica numbers attached to
the mode $m''$\footnote{It is a single number $(\mu)$ if $m''$ is an
anomalous mode, whereas replicon modes are labeled by a pair of
replicas, as explained in \appendixname\ A, $(\mu\nu)$ or $(\mu\bar\mu)$.
There is, of course, no restriction if $m''$ is the longitudinal mode.}
must occur either in $m$ or in $m'$; otherwise the vertex is zero.

Hereinafter we give a list of the nonzero vertices.
To confine the extent of the paper, vertices with
replicon modes of type-II will also be omitted, although they are
available; these vertices are necessary only for a calculation higher order
than one-loop. The presentation follows the order introduced in
Eq.~(\ref{canonicalform}), different symbols are used to indicate
different replicas.

\subsection*{RRR:}

\setlength{\parindent}{0pt}

\newcommand{\vertex}[4]{\mbox{$
\begin{gathered}[c]
\begin{picture}(20,40)(-10,-20)
\put(-10,10){\vector(1,-1){10}}\put(-10,-10){\vector(1,1){10}}
\put(0,0){\vector(1,0){14}}
\put(-4,6){\makebox(0,0)[bl]{$\scriptscriptstyle #1$}}
\put(-4,-9){\makebox(0,0)[bl]{$\scriptscriptstyle #2$}}
\put(9,2){\makebox(0,0)[bl]{$\scriptscriptstyle #3$}}
\end{picture}\end{gathered}
\begin{gathered}[c]{} =#4\end{gathered}$}}
\vertex{\mu\nu}{\omega\rho}{\mu\nu}{-2g_1+g_2}
\hfill
\vertex{\mu\nu}{\omega\rho}{\mu\omega}{
\frac{n-1}{2(n-2)}(-ng_1+g_2)}

\begin{center}\vertex{\mu\omega}{\nu\omega}{\mu\nu}{\frac{n-1}{2(n-2)}
\big[(n-1)(n-4)g_1+g_2\big]}\end{center}

\begin{center}\vertex{\mu\omega}{\nu\omega}{\mu\omega}{\frac{1}{2(n-2)}
\big[(n^2-9n+12)g_1-(n^2-6n+7)g_2\big]}\end{center}

\begin{center}\vertex{\mu\nu}{\mu\nu}{\mu\nu}{\frac{1}{2(n-2)}
\big[2(3n^2-15n+16)g_1+(n^3-9n^2+26n-22)g_2\big]}\end{center}

\subsection*{RRA:}

\begin{center}\vertex{\mu\nu}{\omega\rho}{\mu}{-\frac{2(n-1)^2}{n(n-2)}\,g_3}
\end{center}

\begin{center}\vertex{\mu\omega}{\nu\omega}{\omega}{}
\vertex{\mu\omega}{\nu\omega}{\mu}{\frac{(n-1)^2(n-4)}{n(n-2)}\,g_3}
\end{center}

\begin{center}\vertex{\mu\nu}{\mu\nu}{\mu}{-\frac{(n-1)^2
(n-3)(n-4)}{n(n-2)}\,g_3}
\end{center}

\hspace{4em}\hrulefill\hspace{4em}

\vertex{\mu\nu}{\omega}{\mu\nu}{2g_3}\hfill
\vertex{\mu\nu}{\omega}{\mu\omega}{\frac{n}{n-2}\,g_3}

\begin{center}\vertex{\mu\nu}{\mu}{\mu\nu}{-\frac{(n-1)(n-4)}{n-2}\,g_3}
\end{center}

\subsection*{RRL:}

\vertex{\mu\nu}{\omega\rho}{\text{L}}{\frac{2(n-2)}{n}\,g_4}
\hfill
\vertex{\mu\omega}{\nu\omega}{\text{L}}{-\frac{(n-2)(n-3)}{n}\,g_4}

\begin{center}\vertex{\mu\nu}{\mu\nu}{\text{L}}{\frac{(n-2)^2(n-3)}{n}\,g_4}
\end{center}

\hspace{4em}\hrulefill\hspace{4em}

\begin{center}\vertex{\mu\nu}{\text{L}}{\mu\nu}{2g_4}
\end{center}

\subsection*{RAA:}

\vertex{\mu\nu}{\omega}{\omega}{-\frac{4}{n-2}\,g_5}\hfill
\vertex{\mu\nu}{\omega}{\mu}{-\frac{2(n-1)}{n-2}\,g_5}

\vertex{\mu\nu}{\mu}{\nu}{\frac{2(n-1)(n-3)}{n-2}\,g_5}\hfill
\vertex{\mu\nu}{\mu}{\mu}{\frac{2(n-3)}{n-2}\,g_5}

\hspace{4em}\hrulefill\hspace{4em}

\begin{center}\vertex{\mu}{\nu}{\mu\nu}{\frac{2n^2}{(n-1)(n-2)}\,g_5}
\end{center}

\subsection*{AAA:}

\vertex{\mu}{\nu}{\mu}{\frac{4}{n-2}\,g_6}\hfill
\vertex{\mu}{\mu}{\mu}{-4g_6}

\subsection*{AAL:}

\vertex{\mu}{\nu}{\text{L}}{-\frac{2}{n-1}\,g_7}\hfill
\vertex{\mu}{\mu}{\text{L}}{2g_7}

\hspace{4em}\hrulefill\hspace{4em}

\begin{center}\vertex{\mu}{\text{L}}{\mu}{\frac{4}{n-2}\,g_7}
\end{center}

\subsection*{LLL:}

\begin{center}\vertex{\text{L}}{\text{L}}{\text{L}}
{\frac{2}{n(n-1)}\,g_8}\end{center}

\end{document}